# Models of Consensus for Multiple Agent Systems

Daniel E. O'Leary
3660 Trousdale Parkway
University of Southern California
Los Angeles, CA 90089-1421
oleary@rcf.usc.edu

## Abstract

Models of consensus are used to manage multiple agent systems in order to choose between different recommendations provided by the system. It is assumed that there is a central agent that solicits recommendations or plans from other agents. That agent then determines the consensus of the other agents, and chooses the resultant consensus recommendation or plan. Voting schemes such as this have been used in a variety of domains, including air traffic control.

This paper uses an analytic model to study the use of consensus in multiple agent systems. The binomial model is used to study the probability that the consensus judgment is correct or incorrect. That basic model is extended to account for both different levels of agent competence and unequal prior odds. The analysis of that model is critical in the investigation of multiple agent systems, since the model leads us to conclude that in some cases consensus judgment is not appropriate. In addition, the results allow us to determine how many agents should be used to develop consensus decisions, which agents should be used to develop consensus decisions and under which conditions the consensus model should be used.

## 1 INTRODUCTION

A primary concern of multiple agent systems is the co-ordination of behavior among a collection of intelligent agents. Those agents may have different knowledge, different information, different information sources or all three. The agents may be different algorithms or use different solution processes. As a result, their recommendations or solutions may differ, creating a management task of choosing between potentially conflicting solutions.

This paper considers those systems where there is task centralization. In one version of such systems, there are a set of agents that all solve the same problem. Then another agent is charged with soliciting recommendations or plans from each of the other agents and determining the consensus solution. Thus, the solution must be voted on or chosen from those proposed by the agents. For example, in some mission critical situations, three computer systems are each used to solve the same problem. At some point the three solutions are examined and compared, and the solution that appears in two or more of the three systems is chosen as the correct solution. This is the so-called "majority votes" or consensus situation.

Voting schemes have been used in the development of multiple agent systems. For example, in a series of papers on air traffic control (e.g., Steeb et al. 1981, McArthur et al. 1982 and Cammarata et al. 1983), voting-based schemes were used to control interacting independent agents using task centralization.

The focus of this paper is that of determining under what conditions consensus is likely to generate the correct decision. Thus, the purpose of this paper is to develop and explore analytic models of consensus in order to (a) study conditions under which consensus is a reasonable basis of resolving conflicting decisions, (b) to structure the use of consensus as a multiple agent management tool in information systems. In so doing, this paper provides theoretic foundations of consensus, and a basis for the use of consensus in multiple agent systems. In so doing, this paper provides theoretic foundations for the use of consensus, and a basis for the use of consensus in multiple agent systems.

### 1.1 CONSENSUS AS A BASIS OF COMPARISON

Generally, a lack of consensus among a group of agents implies that some of the agents are not correct. However, even complete agreement does not guarantee correctness. Thus, Einhorn (1974) argued that consensus is a necessary, but not sufficient condition for correctness of agent judgment.

Unfortunately, there is little evidence on the relationship between consensus and correctness. Empirically, researchers have found that individual experts in some domains have been correct only 45 to 85% of the time (e.g., Libby 1976). As noted by Johnson (1988), experts have problems in those domains where there is an uncertain relationship between inputs and outputs, commonly referred to as those situations where there is decision under uncertainty. In those domains, consensus generally is found to provide a good solution. For example, Libby (1976) found that the consensus judgment was correct more often than 42 out of 43 different individuals. Thus, there is interest in determining some of the characteristics of consensus, e.g., under what conditions should we expect the probability of consensus to be greater than a non consensus decision.

Previous researchers (e.g., Cammarata et al. 1983, p. 770) have argued that "Potentially, a group of agents should be able to solve problems more effectively than the same agents working individually. In practice, however, groups often work ineffectively and their joint productivity is less than the sum of the productivities expected of each member." This suggests that in some cases, approaches, such as consensus, do not work as well as might be anticipated, a priori. As a result, this paper is concerned with the determination of such situations, in order to guide system development.

## 1.2 THIS PAPER

This paper proceeds as follows. Section 2 develops a basic model of the correctness of consensus judgments. That section summarizes some classic research as applied to the consensus problem. Section 3 investigates some extensions of that model, by relaxing assumptions inherent in that model. Some of the results of section 3 are new, such as the conditions for use of consensus in the situation of unequal prior odds. Section 4 reviews some of the implications of these models and briefly discusses issues in their implementation. Section 5 provides a brief summary and some extensions.

## 2 A CONSTANT PROBABILITY MODEL

Throughout this paper, it is assumed that there is a single coordinating agent that is ultimately concerned with dichotomous decisions or recommendations. That agent polls the other individual agents in order to choose between two alternatives over a series of decision situations. For example, the system may be asked to decide whether loan applicants will default or not default, or, should a plane land or not land, at this time. The coordinating agent is responsible for choosing the consensus situation, feeding that solution back to the agents and possibly acting on the plan (e.g., Cammarata et al. 1983).

The multiple agent system is assumed to employ one coordinating agent. In addition there are n other agents, each with an equal probability of being correct. Since consensus is the concern in this paper, those n agents are the focus of the remainder of the paper. It is assumed that the probability of success is constant for each problem and that the agents are assumed to arrive at their decisions independently. The agents' decisions are then summarized to determine the consensus judgment.

### 2.1 BACKGROUND

Condorcet (1785) first recognized that Bernoulli's (1713) work on the binomial could be used to model the probability of reaching correct decisions under different voting systems. Condorcet's (1785) work become the basis of modern research in voting (e.g., Black 1958) and jury decision making (Gorfman and Owen 1986). One of the common themes of that research is to determine the probability that the consensus position is correct.

The binomial consists of n independent trials, where each dichotomous choice decision (referred to as the first and second alternatives), has a probability p of success and a probability (1-p) of failure. In a multiple agent setting, the use of the binomial would assume that each of the agent swould have equal competence. In addition, it would also be assumed that each of the two alternatives were equally likely to be correct. This assumption of equal prior odds creates a special case of the binomial, analogous to the case of using a fair coin. Both of these assumptions will be relaxed later in the paper.

### 2.2 A MODEL OF THE CONSENSUS-CORRECTNESS RELATIONSHIP

Let n be the number of agents. Let M be the minimum number of agents necessary to establish a majority. When n is odd, $M=(n+1)/2$, when n is even $M=(n/2)+1$. Let m be the number of agents for a given consensus, where $m = M, ..., n$. Let $P_C$ be the probability of consensus.

Given the two assumptions from the previous section,

$$P_C = \sum_{m=M}^{n} \binom{n}{m} p^m (1-p)^{n-m} .$$

A set of binomial table values for $P_C$ for some values of p and n is given in Table 1.

### 2.3 SOME RESULTS FROM THE MODEL

Condorcet (1785) found a number of important results from the use of the binomial as a model of consensus. Assume that n is odd and n > 3 (although the results could be extended to even sets of agents).

Table 1
Probability of Consensus Being Correct
Assumes Equal Prior Odds

| n | p=.10 | p=.30 | p=.50 | p=.70 | p=.90 |
|---|---|---|---|---|---|
| 3 | .028 | .216 | .500 | .784 | .972 |
| 5 | .009 | .163 | .500 | .837 | .991 |
| 7 | .003 | .126 | .500 | .874 | .997 |
| 9 | .001 | .099 | .500 | .901 | .999 |

Result 1

If $p > .5$ then $P_C > p$.

Result 2

If $p > .5$ then $P_C$ is monotonically increasing in n with a limit of 1.

Result 3

If $p = .5$ then $P_C = .5$.

Result 4

If $p < .5$ then $P_C$ is monotonically decreasing in n with a limit of 0.

Result 5

If $p < .5$ then $P_C < p$.

Result 1 indicates that if $p > .5$ then the probability that the consensus decision is correct, is greater than the probability that any single agent's decision is correct. In this situation, consensus is an appropriate surrogate for correctness.

Result 2 suggests that, if $p > .5$ then the larger the number of agents, the higher that the probability of consensus is correct. This suggests that we have systems with larger number of agents in these situations.

Result 3 finds that in this specific case nothing is gained by going from individual judgments to consensus judgments. If the probability of agents being correct is .5 then the probability that consensus of agents is correct is also .5.

Result 4 indicates that, if $p < .5$ then the larger the number of agents, the lower that the probability of consensus is correct. In this situation, we would not gain from the use of more agents.

Finally, Result 5 finds that if $p < .5$ then the probability that the consensus decision is correct, is less than the probability that a single decision is correct. In this situation, consensus actually results in a lower probability of correctness.

Thus, unless $p > .5$ consensus is not an appropriate management strategy of multiple agents. In addition, if $p > .5$ then the larger the set of agents the higher the probability that the consensus decision will be correct.

## 3 EXTENSIONS OF THE BASIC MODEL

There were two primary assumptions in the model of the previous section: equal competence of agents and equal prior odds. This section extends the model of the previous section by relaxing these assumptions.

### 3.1 RELAXATION OF THE EQUAL COMPETENCE ASSUMPTION

It is reasonable to assume that different agents will have a different probability of providing the correct decision, particularly if they have either different knowledge or information. For example, human experts are often delineated as having different titles indicating gradation in expertise. Thus, it is reasonable to assume that the agents come from a number of different classes, where within each class, each agent is equally competent, yet there is an ordering of the competence of the different classes.

Assume there are two different groups of agents, A and B (this assumption could be extended to more than two groups). It is assumed that within either of those two groups the quality of decisions is equal. Let $p_i$ be the probability that an individual agent in group i is correct, $i = A$ or B. Assume that $.5 < p_A < 1$ and that $p_B < p_A$. Let $P_{C(i)}$ be the probability that a consensus decision of group i is correct, $i = A$, B or, A and B (written as A,B). Margolis (1976) examined the model with this revised assumption and developed the following three results.

Result 6

If $p_b < .5$ then $P_{C(A,B)} < P_{C(A)}$.

Result 7

If $p_b > .5$, then there exists some cardinality of group B, referred to as a critical value B*, such that $P_{C(A,B)} > P_{C(A)}$.

Result 8

There exists some value $p_{b^*} < p_a$, such that if $p_b > p_{b^*}$ then $P_{C(A,B)} > P_{C(A)}$.

Result 6 indicates that if the value of $p_b$ is low enough then it does not make sense to aggregate the two classes of agents in order to develop a consensus value. Result 7 indicates that for $p_b$ of an appropriate level, if group B is large enough then it makes sense to integrate the agents into one large group of A and B, that will make the consensus decision. Result 8 indicates that if $p_b$ is large enough then group B should be integrated with group A, regardless of the size of group B. These results are surprising to a certain extent, since they indicate that, in some situations, lower quality agents should be integrated with higher quality agents for consensus judgments.

Result 7 may lead to the requirement that group B be quite large, so as to be impractical in the case of multiple agent systems. If there are 30 agents in A, $p_a = .7$ and $p_b = .51$ then B* would be several hundred, and thus beyond the scope of typical multiple agent systems.

Using results from Margolis (1976), the critical point nature of Result 8 can be exemplified as follows. If $p_a = .9$ then $p_{b^*} = .82$. If $p_a = .8$ then $p_{b^*} = .70$. If $p_a = .7$ then $p_{b^*} = .62$. If $p_a = .6$ then $p_{b^*} = .55$.

These results can be extended. For example, the following result indicates that if a subset of some set of agents is being used to develop a consensus judgment, then it is always better to add more of those same equal agents to the set of agents from which consensus is being developed.

Result 9

Let A* be a subset of A. $P_{C(A)} > P_{C(A^*)}$ for all A*, not equal to A.

### 3.1.1 Normal Approximation

The normal distribution can be used as an approximation to the binomial (Feller 1950). Thus, an alternative approach has been developed by Grofman (1978) and Grofman et al. (1983) that employs this result. Rather than multiple distinct sets of agents, they treat the set of agents as a single class, with competency normally distributed with a mean of p# and a variance of p#(1-p#)/n. In that case, the conclusions of the equal competence model will hold, with p# substituting for p.

### 3.1.1 Poisson Approximation

The poisson distribution also can be used to approximate the binomial (Feller [1950]), where the poisson is defined as $p(k;L) = e^{-L} * L^k / k!$. In the same sense that the normal approximation to the binomial can be used to develop an alternative approach to the multiple classes, so can the poisson distribution. In the approximation of the poisson distribution, the parameter L is equal to $n^*(1-p)$. With L specified as $n^*(1-p)$ the same results as in section 2 hold. the only constraint on L is that L reflects the density of correct judgments in the group of agents.

### 3.2 RELAXATION OF THE EQUAL PRIOR ODDS ASSUMPTION

The model in section 2 also assumes that there are equal prior odds as to which of the alternatives is correct. However, in most decision making situations it is unlikely that the relevant states of nature are equally likely.

Let $p_S$ be the probability of the first state of the dichotomous decision occurring. Let $p_{S'} = (1 - p_S)$, be the probability of the other state of nature. In the case of equal prior odds, $p_{S'} = p_S = .5$. Let pR be the probability of the agent making the correct decision in favor of the first alternative, given the prior odds for the state of nature S, assuming all agents are of equal competence. Let pR' be the probability of the agent choosing alternative R', making the correct decision, given the prior odds for the state of nature S' and assuming equal competence.

Table 2
Probability of an Individual Decision Being Correct Given Unequal Prior Odds and Various Competencies When Prior Odds are Equal

| Prior Odds | Competency (p) for Equal Prior Odds | | | | |
|---|---|---|---|---|---|
| | p=.10 | p=.30 | p=.50 | p=.70 | p=.90 |
| .10 | .012 | .045 | .100 | .206 | .500 |
| .20 | .027 | .097 | .200 | .368 | .692 |
| .30 | .045 | .155 | .300 | .500 | .794 |
| .40 | .069 | .222 | .400 | .609 | .857 |
| .50 | .100 | .300 | .500 | .700 | .900 |
| .60 | .143 | .391 | .600 | .778 | .931 |
| .70 | .206 | .500 | .700 | .845 | .955 |
| .80 | .308 | .632 | .800 | .903 | .973 |
| .90 | .500 | .794 | .900 | .955 | .988 |

Using Bayes' theorem, we have $p_R = (p*p_S)/[(p*p_S) + (1-p)*p_{S'}]$ and $p_{R'} = (p*p_{S'})/[(p*p_{S'}) + (1-p)*p_S]$. Some example values are given in Table 2.

### 3.2.1 Relationship Between p and $p_S$, and $p_R$

There are a number of relationships between p, $p_S$ and $p_R$, that can be developed, mapping the revised model into results obtained for the basic model, discussed in results 1-5.

Result 10

If $p + p_S > 1$ then $p_R > .5$.

Proof of Result 10

$$p_R = (p*p_S) / [p*p_S + (1-p)*(1-p_S)]$$

$$p_R = (p*p_S) / [2p*p_S + (1 - p - p_S)]$$

Since $(1 - p - p_S)$ is less than 0, $p_R > .5$

Result 11

If $p + p_S < 1$ then $p_R < .5$.

Result 12

If $p + p_S = 1$ then $p_R = .5$.

Proof of Result 12

$$p_R = (p*p_S) / [p*p_S + (1-p)*(1-p_S)]$$

$$p_R = (p*p_S) / [2p*p_S + (1 - p - p_S)]$$

Since $p + p_S = 1$, $p_R = .5$

How is the probability that a consensus judgment is correct impacted by the unequal prior odds? Results 1-5 when combined with Results 10-12 provide us with the answer. We should use consensus only if $p + p_S > 1$. Thus, the quality of consensus judgments is a function of both those probabilities.

### 3.2.2 Monotonicity Result for Revised Model

In addition, we can establish a monotonicity result for $p_R$. In particular, the following result indicates that $p_R$ is monotonically increasing as the prior odds increase.

Result 13 $p_R$ is monotonically increasing in $p_S$.

Proof

Let $p_S > p_S''$, then

$$(p*pS) / [p*p_S + (1-p)*(1-p_S)] >$$

$$(p*p_S'') / [p*pS'' + (1-p)*(1-p_S'')]$$

$$(p*pS) [2p*p_S'' + 1 - p - p_S''] >$$

$$(p*pS'') [2p*p_S + 1 - p - p_S]$$

$$p*p_S - p*p*p_S > p*pS'' - p*p*pS''$$

$$p*p_S - p*p_S'' > p*p*p_S - p*p*p_S''$$

Since $p_S > p_S''$, the inequality holds and $p_R$ is monotonically increasing in $p_S$.

This can be a useful result. For example, we can make the following two statements. First, if we know p and have a conservative estimate of $p_S$, such that $p + p_S > 1$, then we know that we should use consensus. We do not need to know $p_S$ exactly. We may be able to use simply a lower bound. Second, if the prior odds are greater than .5, we know that the simplified equal odds model underestimates $p_R$.

Thus, in some cases the equal prior odds model helps bound the case where the prior odds are not equal.

## 4 IMPLICATIONS AND IMPLEMENTATION

This section discusses some of the implications of the models in this paper and their implementation.

### 4.1 IMPLICATIONS

The basic model and its extensions, discussed in sections 2 and 3, has a number of implications. First, the model indicates that the decision on whether or not consensus should be used to manage multiple agent systems, is a function of the sum of two parameters: p and $p_S$. Consensus should not be used indiscriminately. Second, in the consensus decision in the basic model, where $p > .5$, it is always beneficial for the use of a complete set of the best agents. If all the top agents cannot participate then it is likely that the next highest class of agents should also be used in the development of the consensus judgment. Third, in the consensus model where $p > .5$ the results here suggests developing systems that have as large a set of agents as is feasible. At the margin, development costs can be traded off against the ability of the next agent to improve the probability that the consensus judgement is correct. Fourth, the models imply some stopping criteria in the design of multiple agent systems. For example, we can see from table 1, that if there are equal prior odds and if $p = .70$ and we wish a $p_C > .8$ then we must use at least 5 agents to develop the consensus judgment.

### 4.2 SOME IMPLEMENTATION CONSIDERATIONS

In order to implement the models in this paper, basic knowledge of the underlying parameters is required. The probabilities p and $p_S$, are necessary to use the binomial model. The competency levels p could be obtained using at least two different approaches.

First, a set of experiments could be generated to determine the probability of correct judgment (e.g., Libby 1976). Second, past performance data could be used. Prior odds of events, $p_S$ could be obtained from experience. However, there is little in the literature about the quality of competence in even broad categories of events. This is an area for future research.

## 5 SUMMARY AND EXTENSIONS

This paper provides an analysis of a model of consensus for investigation of multiple agent problems with task centralization. There is an agent that solicits recommendations or plans from n independent agents. That centralized agent is then responsible for determining the consensus of the n agents.

Consensus models have been used in a number of domains, including mission critical situations and air traffic control models. The consensus model is based on the binomial, but was extended to include multiple levels of competence and unequal prior odds. The results presented here summarized some classic results and presented new results.

The basic model was limited to simple majorities as the means of the definition of consensus. Alternative approaches used by other organizations may include a two-thirds majority. These alternative definitions of consensus could be accounted for in the model developed above. Further, rather than binomial models, multinomial models could be developed. Another approach is to use a Bayesian model of consensus.

For example, O'Leary (1994) has studied the impact of using a Bayesian model, assuming changes in probabilities of individual agents on sequential judgment situations, based on sequential successes and failures of the agents to develop correct solutions.

Further, the consensus approach could be compared to a complete decision analytic approach. In this setting, each agent would be viewed as a noisy sensor, mapping into a set of binary outputs. The problem would then be to choose different policies in order to maximize the expected utility, given a set of inputs. Voting policies, such as consensus, could provide efficient approximations to complete decision theoretic approaches.

### Acknowledgements

The author would like to acknowledge discussions with Professor K. Pincus on consensus. In addition, the author would like to acknowledge the extensive comments by the three referees on an earlier version of this paper. Their comments helped make this a better paper.

### References

J. Bernoulli, (1713). Ars Conjectandi (The Art of Conjecturing), first published posthumously in Latin (Later published as Wahrscheinlichkeitsrechnung, Leipzig, Engelmann, 1899).

D. Black, (1958). The Theory of Committees and Elections London: Cambridge University Press.

A. Bond and L. Gasser (1988). Readings in Distributed Artificial Intelligence, San Mateo, CA: Morgan Kaufman.

S. Cammarata, D. McArthur, and R. Steeb, (1983). "Strategies of Cooperation in Distributed Problem Solving," Proceedings of the 1983 International Joint Conference on Artificial Intelligence: 767-770.

Condorcet, M. [Marie Jean Antoine Nicolas Caritat, Marquis de Condorcet] (1785). Essai sur l'application de l'analyse a la probabilitie des voix, [Essay on the Application of Analysis to the Probability of Majority Decisions], Paris, Imprimerie Royale.

H. Einhorn, (1974). "Expert Judgment: Some Necessary Conditions and an Example," Journal of Applied Psychology, 59 (5): 562-571.

W. Feller, (1950). An Introduction to Probability Theory and its Applications, London: John Wiley.

L. Goldberg and C. Werts, (1966). "The Reliability of Clinician's Judgments: A Multi-Method Approach," Journal of Consulting Psychology, June, 199-206.

B. Grofman, (1978). "Judgemental Competence of Individuals and Groups in a Dichotomous Choice Situation," Journal of Mathematical Sociology, 6: 47-60.

B. Grofman and G. Owen, (1986). "Condorcet Models, Avenues for Future Research," in Information Pooling and Group Decision-Making: Proceedings of the Second University of California, Irvine, Conference on Political Economy, B. Grofman and G. Owen (ed)

B. Grofman, G. Owen, and S. Feld, (1984). "Thirteen Theorems in Search of the Truth," Organizational Behavior and Human Performance, 33: 350-359.

E. Johnson, (1988). "Expertise and Decision Under Uncertainty: Performance and Process," 209-228 in The Nature of Expertise, M. Chi, R. Glaser, and M. Farr, Hillsdale, NJ: LEA.

R. Libby, (1976). "Man versus Models of Man: Some Conflicting Evidence," Organizational Behavior and Human Performance, 16 (1): 1-12.

H. Margolis, (1976). "A Note on Incompetence," Public Choice, 26: 119-127.

D. McArthur, R. Steeb, S. Cammarata, (1982). "A Framework for Distributed Problem Solving," Proceedings of the National Conference on Artificial Intelligence, Pittsburg: 181-184.

D. O'Leary, (1994). "Bayesian Models of Consensus for Multiple Agent Systems," unpublished working paper, University of Southern California, January.

R. Steeb, S. Cammarata, F. Hayes-Roth, P. Thorndyke, and R. Wesson, (1981). "Architectures for Distributed Intelligence for Air Fleet Control," TR R-2738-ARPA, Rand Corporation, reprinted in part in Bond and Gasser [1988]: 90-101.